\documentclass[reprint,aps,prd,onecolumn,notitlepage,nofootinbib,preprintnumbers,superscriptaddress]{revtex4}
\usepackage{amsmath}
\usepackage{amsfonts}
\usepackage{amssymb}
\usepackage{latexsym}
\usepackage{enumerate}
\usepackage{color,xcolor}
\usepackage{graphicx}
\usepackage{bm}
\usepackage{epsfig}
\usepackage[english]{babel}
\usepackage{times}
\usepackage{comment}
\usepackage{hyperref}
\usepackage{epstopdf}
\usepackage{extarrows}

\def\bA{\mathbf{A}}
\def\bE{\mathbf{E}}
\def\bB{\mathbf{B}}

\def\aF{\tilde{F}}
\def\tT{\tilde{T}}
\def\Lag{\mathcal{L}}
\def\d{\mathrm{d}}

\begin{document}
\title{Transverse electric waves in Bandos-Lechner-Sorokin-Townsend nonlinear electrodynamics}
\author{Yang Shi}
\affiliation{School of Physics, East China Normal University, Shanghai 200241, China\\}
\author{Qinyan Tan}
\affiliation{Shanghai Institute of Spaceflight Control Technology, Shanghai 201109, China\\ \vspace{0.2cm}}
\author{Towe Wang}
\email[Electronic address: ]{twang@phy.ecnu.edu.cn}
\affiliation{School of Physics, East China Normal University, Shanghai 200241, China\\}
\date{\today\\ \vspace{1cm}}
\begin{abstract}
In the generalized Born-Infeld electrodynamics discovered by Bandos, Lechner, Sorokin and Townsend, we study transverse electric waves propagating perpendicular to a constant magnetic field background in a parallel-plate waveguide. The directions of propagation and polarization of the waves are perpendicular to each other, and both of them are parallel to the perfectly conducting plates. Two specific configurations are studied, in which the background magnetic field is either normal to the plates or along the polarization direction. The dispersion relation, the velocity and the cutoff frequency of the lowest-order lowest-frequency mode are calculated in both configurations. This paves the way for a potential test of the generalized Born-Infeld electrodynamics.
\end{abstract}

\maketitle
\section{Introduction}\label{sect-intro}
Maxwell electrodynamics is the foundation not only of electrical engineering but also of modern physics. Its invariance across different reference frames has led to the theory of special relativity, in which the Lagrangian of a point mass is $L_\mathrm{rel}=mc^2(1-\sqrt{1-v^2/c^2})$ with an upper bound of velocity. In analogy to this Lagrangian, Born and Infeld developed a theory with the Lagrangian density
\begin{equation}\label{LBI}
\Lag_\mathrm{BI}=T\left(1-\sqrt{1-\frac{|\bE|^2-|\bB|^2}{T}}\right)
\end{equation}
which puts a finite upper bound to the self-energy of a point charge \cite{Born:1934gh}. Here and below $T$ is a constant parameter, and we refer to it as Born-Infeld constant. Afterwards nonlinear electrodynamics of various forms were constructed as fundamental or effective theories or toy models and revived from time to time \cite{Heisenberg:1936nmg,BialynickiBirula:1984tx,Bialynicki-Birula:1992rcm,Soleng:1995kn,Gaete:2013dta,Jing:2011vz,Kruglov:2014hpa,Bandos:2020jsw,Kosyakov:2020wxv,Gullu:2020ant,Russo:2025fuc}, see Refs. \cite{Dunne:2012vv,Sorokin:2021tge} for more comprehensive reviews of literature.

As is well known, Maxwell electrodynamics is invariant under both conformal transformations and $SO(2)$ electromagnetic duality transformations \cite{Jackson:1999ny,Sorokin:2021tge}. It is less known that Bialynicki-Birula electrodynamics \cite{BialynickiBirula:1984tx}, which was first found as a strong-field limit of Born-Infeld electrodynamics, is invariant under both conformal transformations and $SL(2,R)$ electromagnetic duality transformations \cite{Bialynicki-Birula:1992rcm}. Until several years ago, it was unknown that there is a nonlinear duality-invariant conformal extension of Maxwell electrodynamics, dubbed then as Modified Maxwell (ModMax) electrodynamics \cite{Bandos:2020jsw}.

ModMax electrodynamics was discovered by Bandos, Lechner, Sorokin and Townsend as a weak-field limit of generalized Born-Infeld electrodynamics in Ref. \cite{Bandos:2020jsw}, where they also proved that the only conformal and duality invariant electrodynamics theories are Bialynicki-Birula electrodynamics and ModMax electrodynamics. From then on, considerable efforts have been made to study ModMax electrodynamics, but less attention has been paid to the generalized Born-Infeld electrodynamics, to which we will refer as Bandos-Lechner-Sorokin-Townsend (BLST or BLeST) electrodynamics. As will be explained below, BLST electrodynamics is a remarkable theory that unifies Maxwell, Born-Infeld, Bialynicki-Birula and ModMax theories.

Like Born-Infeld electrodynamics, BLST electrodynamics preserves the $SO(2)$ electromagnetic duality invariance but breaks the conformal invariance. Its Lagrangian density \cite{Sorokin:2021tge} can be written as
\begin{equation}\label{LBLST}
\Lag_\mathrm{BLST}=T-\sqrt{T^2-2T\Lag_\mathrm{ModMax}-\frac{1}{16}\left(F_{\mu\nu}\aF^{\mu\nu}\right)^2}
\end{equation}
in terms of the electromagnetic field strength $F_{\mu\nu}$ and its dual $\aF^{\mu\nu}$, where the Lagrangian density of ModMax electrodynamics
\begin{equation}\label{LModMax}
\Lag_\mathrm{ModMax}=-\frac{\cosh\gamma}{4}F_{\mu\nu}F^{\mu\nu}+\frac{\sinh\gamma}{4}\sqrt{\left(F_{\mu\nu}F^{\mu\nu}\right)^2+\left(F_{\mu\nu}\aF^{\mu\nu}\right)^2}.
\end{equation}
Here $\gamma$ is a constant parameter, which will be referred to as ModMax constant. It is easy to see BLST Lagrangian reduces to Born-Infeld Lagrangian in the special case $\gamma=0$, and to ModMax Lagrangian in the weak-field limit $T\rightarrow\infty$. Furthermore, Maxwell Lagrangian can be recovered from Born-Infeld Lagrangian by taking the limit $T\rightarrow\infty$, or from ModMax Lagrangian by taking the limit $\gamma\rightarrow0$. Just like Born-Infeld theory, in the strong-field limit $T\rightarrow0$, BLST theory reduces to Bialynicki-Birula electrodynamics, as demonstrated in the Hamiltonian formulation in Refs. \cite{BialynickiBirula:1984tx,Russo:2022qvz,Mezincescu:2023zny}.

In the presence of strong external magnetic fields, nonlinear electrodynamics can leave a fingerprint on the electromagnetic waves with distinct dispersion relations. This has been investigated extensively in the literature. To study photon splitting processes, Ref. \cite{Bialynicka-Birula:1970nlh} calculated the group and the phase velocities of electromagnetic waves propagating in a strong magnetic field, while Ref. \cite{Adler:1971wn} calculated the refractive indices of such waves. Further calculations were performed for vacuum dichroism and birefringence effects in a rotating transverse magnetic field by Refs. \cite{Mendonca:2006pg,Adler:2006zs}, in an external plane-wave field by Ref. \cite{Aleksandrov:2021jtj}, and in the nonperturbative high-energy regime by Refs. \cite{Denisov:2016pfu,Bragin:2017yau}. In Ref. \cite{Novello:1999pg}, the propagation of light in the background electromagnetic fields was studied in terms of null geodesics in the effective geometry. This method was later applied to the Born-Infeld theory by Ref. \cite{Aiello:2006cz}, and to the Euler-Heisenberg theory by Refs. \cite{Dinu:2013gaa,Guzman-Herrera:2020ffc}. In a constant magnetic background, Ref. \cite{Neves:2021tbt} investigated the Compton scattering in nonlinear electrodynamic models, such as Hoffmann-Infeld, Euler-Heisenberg, generalized Born-Infeld and Logarithmic models. Recently, related and other phenomena of light propagation in an electric or magnetic background were also discussed in Refs. \cite{Perez-Garcia:2022kvz,Santos:2023ooc} for the Euler-Heisenberg theory, in Refs. \cite{Neves:2022jqq,Cruz:2025xut} for the ModMax theory, and in Refs. \cite{Shi:2024nmx,Shi:2025bri} for the BLST theory, etc.

As far as we know, there have already been some experiments to test predictions of quantum electrodynamics or to search for axionlike particles, and a few of them can be used in a more general way to test nonlinear electrodynamics \cite{Fouche:2016lne,Ellis:2017edi,NiauAkmansoy:2017kbw,NiauAkmansoy:2018ilv,Ejlli:2020yhk,Fedotov:2022ely}. Independent of these experiments and the above-mentioned approaches, in Ref. \cite{Ferraro:2007ec}, Ferraro proposed to test Born-Infeld electrodynamics with waveguides by applying an external magnetic field. Waveguides are fundamental components in radar and communication systems. On the basis of Maxwell electrodynamics, a lot of theoretical works have been done to analyze the electromagnetic waves propagating through them, including some recent works on transverse electric (TE) waves in plane waveguides with corners \cite{Dauge:2012pw} or in plane dielectric waveguides \cite{Smirnov:2015gew,Zhou:2024pte}, and on TE waves in plane waveguides with the control of external magnetic fields \cite{Noble:2008aei,Mohaghegh:2019mpp,Zhang:2020scp,Zhou:2021ewp,Iukhtanov:2023rsw}. Beside these works based on classical electrodynamics, waveguides have also been utilized to study quantum physical properties of atoms strongly interacting with quantized electromagnetic modes \cite{Sheremet:2021krw,Ashida:2021rls}.

\begin{figure}
\centering
\includegraphics[width=0.45\textwidth]{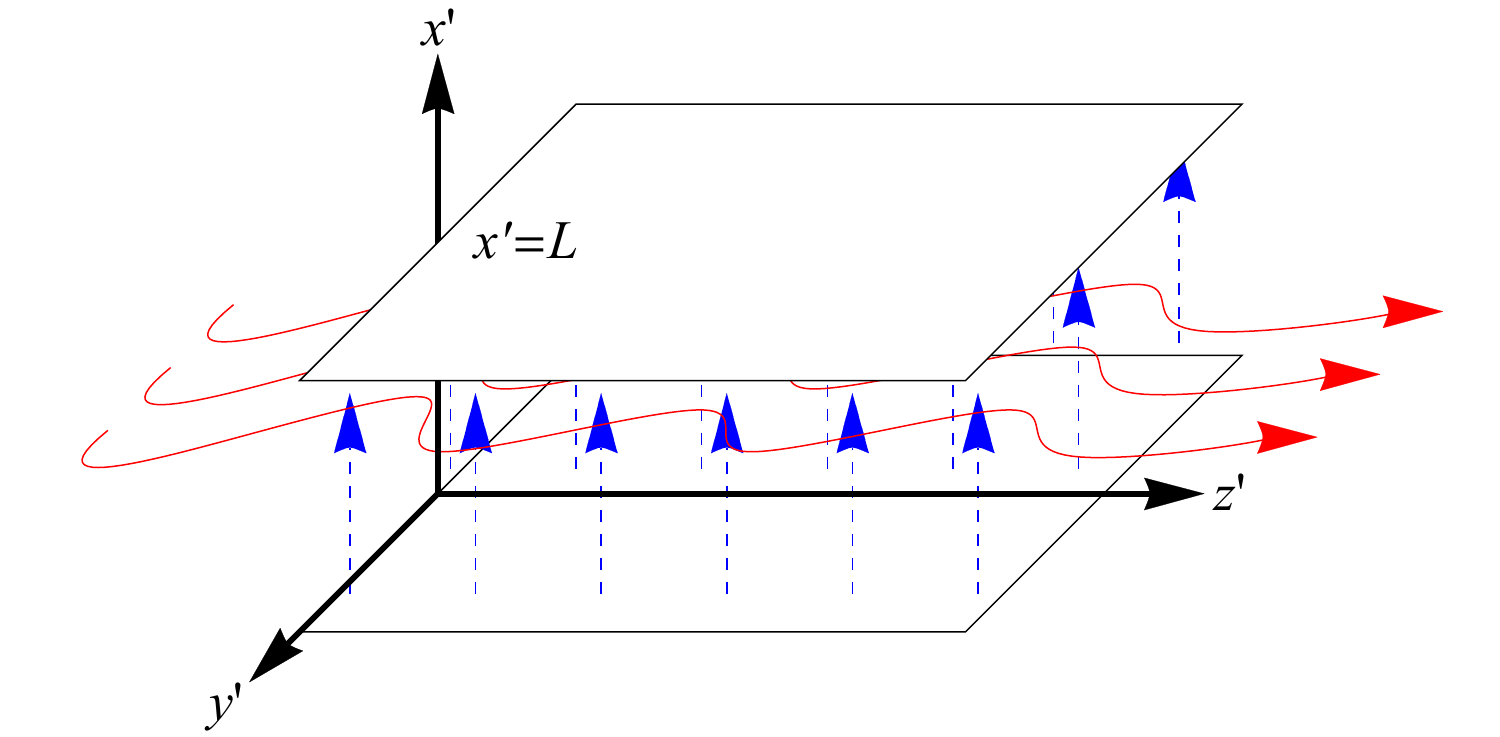}\\
\includegraphics[width=0.45\textwidth]{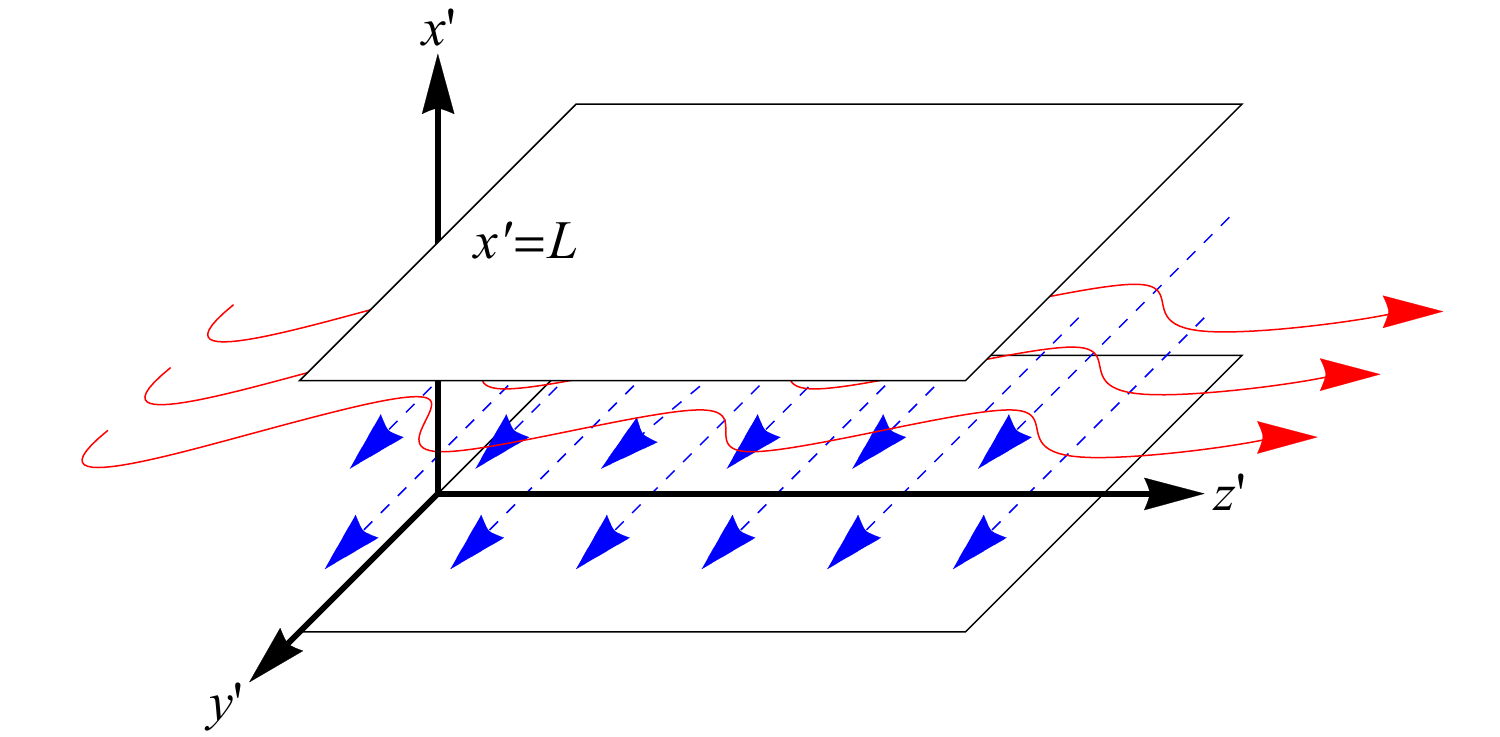}
\caption{Schematic representation of two experimental configurations considered in this article. In both configurations, there are two parallel conducting plates located in the planes $x'=0$ and $x'=L$. Between the plates, the TE waves $u(x',\omega't'-k_{\|}z')$ (red sine waves) are propagating in the $z'$-direction and polarized along the $y'$-axis. The constant background magnetic field $\bB'_0$ (blue dashed lines) is along the $x'$-axis in the upper subfigure, and along the $y'$-direction in the lower subfigure, corresponding to the configuration in Sec. \ref{sect-Bx} and the configuration in Sec. \ref{sect-By}, respectively.}\label{fig-conf}
\end{figure}

As we have explained, BLST electrodynamics is a theoretical framework that can unify Maxwell, Born-Infeld, Bialynicki-Birula and ModMax theories. Different theories reside in different parameter regions. From the experimental point of view, BLST electrodynamics can thus serve as a parameterized framework to test these theoretical models. It is interesting to extend Ferraro's test to BLST electrodynamics so that it can cover more models and get more motivation. To this end, in the present article, we will study TE waves of BLST electrodynamics propagating perpendicular to a constant magnetic field background in a parallel-plate waveguide, enlightened by similar ideas on Born-Infeld electrodynamics \cite{Ferraro:2007ec,Ferraro:2003ia,Manojlovic:2020ndn}. By definition, the directions of propagation and electric field of a TE wave are perpendicular to each other, and both of them are parallel to the perfectly conducting plates. Therefore, it is convenient to choose a Cartesian coordinate system where the TE wave is propagating in the $z$-axis and polarized along the $y$-axis, and the plates are normal to the $x$-axis, see Fig. \ref{fig-conf}. For concreteness, the two plates are located at $x=0$ and $x=L$.

Based on BLST electrodynamics, we will work out the dispersion relation of the lowest-order lowest-frequency TE mode (TE$_1$ mode) in two specific configurations. In the first configuration, the background magnetic field is in the $x$-direction, for which our calculation will be carried out in Sec. \ref{sect-Bx}. In the second configuration, the background magnetic field is along the $y$-direction, of which the details will be given in Sec. \ref{sect-By}. In both configurations, one scheme of Ref. \cite{Ferraro:2007ec} will be employed for calculations. We first study a standing-wave solution of BLST electrodynamics between two parallel conducting plates in the presence of constant external electric and magnetic fields. The external electric field $\bE_0$ is perpendicular to both the $z$-axis and the external magnetic field $\bB_0$. In a convenient gauge \cite{Manojlovic:2020ndn}, the standing waves are oscillating electric and magnetic fields along $y$ and $z$ axes, respectively. Performing a suitable Lorentz boost in the $z$-direction, we can then transform the constant external electric and magnetic fields into a constant magnetic field background $\bB'_0$ parallel to $\bB_0$. Meanwhile, the standing waves are transformed into propagating TE waves \cite{Ferraro:2007ec}. Making use of the results, we give a simple guide to the waveguide test of BLST theory in Sec. \ref{sect-test}, and present concluding remarks in Sec. \ref{sect-con}.

We work in natural units with $\mu_0=c=\hbar=1$ and adopt the metric convention ($+,-,-,-$). Following Ref. \cite{Manojlovic:2020ndn}, we use sign conventions for field components as defined in Ref. \cite{Jackson:1999ny}, Section 11.9 and denote derivatives of $u$ as $u_t=\partial_t u$, $u_x=\partial_x u$, etc. Following Refs. \cite{Bandos:2020jsw,Sorokin:2021tge}, we assume $T>0$ and $\gamma\geq0$ throughout this article.

\section{TE waves polarized perpendicular to a constant magnetic field background}\label{sect-Bx}
\subsection{Standing-wave solution}\label{subsect-Bxstand}
In this section, we study TE waves polarized perpendicular to a constant transverse magnetic field in a parallel-plate waveguide. For this purpose, we start with an external magnetic field in the $x$-direction, in addition to which there is an external electric field and an oscillating electric field along the $y$-direction. The waveguide is taken as two perfectly conducting plates parallel to the $yz$-plane, located at $x=0$ and $x=L$. For this configuration, the simplest solution for the electromagnetic field in a convenient gauge \cite{Manojlovic:2020ndn} has the form
\begin{equation}\label{ABx}
A_y(x,y,z,t)=-E_0t-B_0z+u(x,t),~~~~\Phi=A_x=A_z=0
\end{equation}
with $u$ to be determined by field equations. We then have
\begin{eqnarray}
&&\bE=-\partial_t \bA-\nabla\Phi=(0,E_0-u_t,0),\\
&&\bB=\nabla\times\bA=(B_0,0,u_x),
\end{eqnarray}
and in our sign conventions,
\begin{eqnarray}
\label{FFBx}&&\frac{1}{4}F_{\mu\nu}F^{\mu\nu}=\frac{1}{2}\left(|\bB|^2-|\bE|^2\right)=\frac{1}{2}\left[B_0^2+u_x^2-(E_0-u_t)^2\right],\\
&&\frac{1}{4}F_{\mu\nu}\aF^{\mu\nu}=-\bE\cdot\bB=0.
\end{eqnarray}

It is noteworthy that, as long as $F_{\mu\nu}\aF^{\mu\nu}=0$, the Lagrangian density \eqref{LBLST} reduces to
\begin{eqnarray}\label{LBI}
\nonumber\Lag&=&T-\sqrt{T^2+\frac{T}{2}e^{-\sigma\gamma}F_{\mu\nu}F^{\mu\nu}}\\
&=&e^{-\sigma\gamma}\left(\tT-\sqrt{\tT^2+\frac{\tT}{2}F_{\mu\nu}F^{\mu\nu}}\right),
\end{eqnarray}
where we introduced the notation $\tT=e^{\sigma\gamma}T$ with
\begin{equation}\label{sigma}
\sigma=\left\{\begin{array}{ll}
1, & \hbox{if $F_{\mu\nu}F^{\mu\nu}\geq0$;} \\
-1, & \hbox{if $F_{\mu\nu}F^{\mu\nu}<0$.}
\end{array}\right.
\end{equation}
This result shows the BLST Lagrangian has the same form as the Born-Infeld Lagrangian when $\bE\cdot\bB=0$, with the replacement $T\rightarrow e^{\sigma\gamma}T$.

Inserting Eq. \eqref{FFBx} into Eq. \eqref{LBI} leads to the Lagrangian density
\begin{equation}
\Lag=e^{-\sigma\gamma}\left\{\tT-\sqrt{\tT^2+\tT\left[B_0^2+u_x^2-(E_0-u_t)^2\right]}\right\}.
\end{equation}
The corresponding Euler-Lagrange equation is
\begin{equation}\label{ELBx}
u_{xx}-u_{tt}+\frac{1}{\tT}\left\{\left[B_0^2-\left(E_0-u_t\right)^2\right]u_{xx}-\left(B_0^2+u_x^2\right)u_{tt}-2\left(E_0-u_t\right)u_x u_{xt}\right\}=0.
\end{equation}
In the large $\tT$ limit, this equation can be solved iteratively using the Poincar\'{e}-Lindstedt method. The method has been reviewed in Ref. \cite{Manojlovic:2020ndn}. Imposing the boundary conditions $u(0,t)=u(L,t)=0$ and introducing the orthogonal functions
\begin{eqnarray}
\nonumber s_{nm}(x,t)&=&\sin(nkx+m\omega t)+\sin(nkx-m\omega t)\\
\label{snm}&=&2\sin(nkx)\cos(m\omega t),\\
\nonumber c_{nm}(x,t)&=&\cos(nkx+m\omega t)-\cos(nkx-m\omega t)\\
\label{cnm}&=&2\sin(nkx)\sin(m\omega t)
\end{eqnarray}
with $k=\pi/L$, one can formally write down the solution order by order as
\begin{eqnarray}
u&=&u^{(0)}+u^{(1)}+\cdots+u^{(N)}+\mathcal{O}\left(\tT^{-(N+1)}\right),\\
\omega^2&=&\omega^2_{(0)}+\omega^2_{(1)}+\cdots+\omega^{(N)}+\mathcal{O}\left(\tT^{-(N+1)}\right),
\end{eqnarray}
where $u^{(i)}(x,t)~(i\geq0)$ are expected to be linear combinations of the functions $s_{nm}(x,t)$ with coefficients of order $\tT^{-i}$. In accordance with Ref. \cite{Manojlovic:2020ndn}, we assume there are no $s_{nn}(x,t)$ modes in all high order terms $u^{(i)}(x,t)~(i\geq1)$. The lowest-order solution of Eq. \eqref{ELBx} is
\begin{equation}\label{uom0Bx}
u^{(0)}(x,t)=A\sin(kx)\cos(\omega t)=\frac{A}{2}s_{11}(x,t),~~~~\omega^2_{(0)}=k^2
\end{equation}
which serves as the seed solution. This guarantees that the standard results of Maxwell electrodynamics can be reproduced in the limit $\tT\rightarrow\infty$. Substituting $u=u^{(0)}+u^{(1)}$, $\omega^2=\omega^2_{(0)}+\omega^2_{(1)}$ into Eq. \eqref{ELBx} and keeping the $\mathcal{O}(1/\tT)$ terms, one obtains
\begin{equation}
u^{(1)}_{xx}-u^{(1)}_{tt}+\frac{A}{2}\omega^2_{(1)}s_{11}+\frac{1}{\tT}\left[\frac{A}{2}k^2\left(E_0^2+\frac{A^2k^2}{2}\right)s_{11}+\frac{A^3}{8}k^4\left(s_{31}-s_{13}\right)+A^2k^3E_0\sin(2\omega t)\right]+\mathcal{O}\left(\tT^{-2}\right)=0.
\end{equation}
To meet this equation, the $\mathcal{O}(1/\tT)$ terms should cancel out. It therefore follows that
\begin{eqnarray}
\label{u1Bx}u^{(1)}(x,t)&=&\frac{A^3}{64\tT}k^2\left(s_{31}+s_{13}\right)-\frac{A^2}{4\tT}kE_0\sin(2\omega t),\\
\label{om1Bx}\omega^2_{(1)}&=&-\frac{k^2}{\tT}\left(E_0^2+\frac{A^2k^2}{2}\right).
\end{eqnarray}
Altogether, we have the $\mathcal{O}(1/\tT)$ solution of Eq. \eqref{ELBx},
\begin{eqnarray}
\label{uBx}u&=&\frac{A}{2}s_{11}+\frac{A^3}{64\tT}k^2\left(s_{31}+s_{13}\right)-\frac{A^2}{4\tT}kE_0\sin(2\omega t)+\mathcal{O}\left(\tT^{-2}\right),\\
\label{omBx}\omega^2&=&k^2-\frac{k^2}{\tT}\left(E_0^2+\frac{A^2k^2}{2}\right)+\mathcal{O}\left(\tT^{-2}\right).
\end{eqnarray}
This procedure can be continued to higher orders as demonstrated in Ref. \cite{Manojlovic:2020ndn}.

Here are some remarks about the boundary conditions $u(0,t)=u(L,t)=0$. First, these conditions are equivalent to the statement that any nonsingular solution for $u(x,t)~(0\leq x\leq L)$ can be expanded in orthogonal functions $s_{nm}(x,t)$ and $c_{nm}(x,t)~(n\geq1)$. Second, although the boundary conditions are satisfied by $u^{(0)}(x,t)$, they are violated by the term $\sin(2\omega t)$ in $u^{(1)}(x,t)$ (the last term in Eq. \eqref{u1Bx}). This term is suppressed by $1/\tT$ and can be interpreted as a time-periodic uniform electric field. It is in agreement with the fact that the boundary conditions are violated by the last term in Euler-Lagrange equation \eqref{ELBx}, which is also suppressed by $1/\tT$. One can check this by inserting $u(x,t)\propto\sin(nkx)$ into Eq. \eqref{ELBx}. Third, in practice, electric fields induced by the waveguide walls will partially cancel the external electric field $\bE_0$ in the neighborhoods of the surfaces, therefore Eqs. \eqref{ELBx} and \eqref{uBx} are invalid at $x/L\ll1$ or $(L-x)/L\ll1$. Far away from the boundaries, these equations are nevertheless valid. Since we will be mainly interested in the dispersion relation of the TE$_1$ mode in the waveguide, we can safely disregard the mode of frequency $2\omega$ in our analysis.

\subsection{Propagating TE$_1$ mode}\label{subsect-BxTE1}
In the previous subsection, for a configuration with orthogonal electric and magnetic background fields, we have explored the standing-wave solution of BLST electrodynamics. Since we are interested in TE waves polarized perpendicular to a constant transverse magnetic field in a parallel-plate waveguide, let us apply a Lorentz transformation along the $z$-axis to the standing-wave solution.

The Lorentz boost along the $z$-axis can be realized by a coordinate transformation as follows:
\begin{equation}\label{boost}
x\rightarrow x',~~~~y\rightarrow y',~~~~z\rightarrow\frac{z'-Vt'}{\sqrt{1-V^2}},~~~~t\rightarrow\frac{t'-Vz'}{\sqrt{1-V^2}}.
\end{equation}
Here $(t,x,y,z)$ and $(t',x',y',z')$ can be regarded as coordinate systems attached to two reference frames $S$ and $S'$, respectively. The reference frame $S$ is the one in which we were working in the previous subsection. The reference frame $S'$ is moving with velocity $V$ relative to $S$ in the negative direction of $z$-axis. By such a Lorentz boost, the electromagnetic field \eqref{ABx} is transformed into
\begin{eqnarray}
\nonumber A_{y'}&=&-\frac{E_0-VB_0}{\sqrt{1-V^2}}t'-\frac{B_0-VE_0}{\sqrt{1-V^2}}z'+u\left(x',\frac{t'-Vz'}{\sqrt{1-V^2}}\right),\\
\Phi&=&A_{x'}=A_{z'}=0.
\end{eqnarray}
Remember that we are interested in TE waves polarized perpendicular to a constant external magnetic field. In order to get rid of an external electric field in the new frame $S'$, we preset $E_0=VB_0$ \cite{Jackson:1999ny} and thus have
\begin{equation}
\bE'=(0,-u_{t'},0),~~~~\bB'=\left(B'_0-u_{z'},0,u_{x'}\right),
\end{equation}
where the external magnetic field is $B'_0=B_0\sqrt{1-V^2}$ in the $x'$-direction. Corresponding to this case, the experimental configuration is illustrated in the upper panel of Fig. \ref{fig-conf}.

In the reference frame $S'$, Eqs. \eqref{uBx} and \eqref{omBx} take the form
\begin{eqnarray}
\nonumber\label{u'Bx}u\left(x',\frac{t'-Vz'}{\sqrt{1-V^2}}\right)&=&A\sin(k_{\bot}x')\cos(\omega't'-k_{\|}z')-\frac{A^2}{4\tT}k_{\|}B'_0\sin(2\omega't'-2k_{\|}z')\\
&&+\frac{A^3}{32\tT}k_{\bot}^2\left[\sin(3k_{\bot}x')\cos(\omega't'-k_{\|}z')+\sin(k_{\bot}x')\cos(3\omega't'-3k_{\|}z')\right]+\mathcal{O}\left(\tT^{-2}\right),\\
\label{om'Bx}\omega'^2-k_{\|}^2&=&k_{\bot}^2-\frac{1}{\tT}\left(k_{\|}^2B'^2_0+\frac{A^2k_{\bot}^4}{2}\right)+\mathcal{O}\left(\tT^{-2}\right)
\end{eqnarray}
with the wave four-vector $(\omega',k_{\bot},0,k_{\|})$ dictated by
\begin{equation}\label{4wav}
\omega'=\frac{\omega}{\sqrt{1-V^2}},~~~~k_{\bot}=\frac{\pi}{L},~~~~k_{\|}=\omega'V.
\end{equation}
Clearly, by the Lorentz boost, the solution \eqref{uBx} is turned into propagating waves along the $z$-axis. Especially, the $s_{11}$ mode of standing wave is turned into the TE$_1$ mode of propagating wave, while the time-periodic uniform mode is transformed into a uniform plane wave of frequency $2\omega'$ which can be identified as the TE$_0$ mode. By contrast, in Maxwell electrodynamics, the lowest-order mode of the transverse magnetic field is the TM$_0$ mode, whereas the lowest-order TE mode is TE$_1$ in a parallel-plate waveguide. Because we started with the $s_{11}$ mode as the seed solution in Sec. \ref{subsect-Bxstand}, here the obtained dominant mode is TE$_1$ mode in Eq. \eqref{u'Bx}. As can be derived from the dispersion relation \eqref{om'Bx}, its phase velocity $v_p=\omega'/k_{\|}$ and group velocity $v_g=d\omega'/dk_{\|}$ are
\begin{equation}\label{vpvgBx}
v_p=\frac{1}{V},~~~~v_g=\left(1-\frac{B'^2_0}{\tT}\right)V+\mathcal{O}\left(\tT^{-2}\right),
\end{equation}
where $V=k_{\|}/\omega'$ is
\begin{equation}\label{VBx}
V^2=\left(1-\frac{B'^2_0}{\tT}\right)^{-1}\left[1-\frac{k_{\bot}^2}{\omega'^2}\left(1-\frac{A^2k_{\bot}^2}{2\tT}\right)\right]+\mathcal{O}\left(\tT^{-2}\right).
\end{equation}
This expression is consistent with Ref. \cite{Sorokin:2021tge}, Equation (54) in the limit $L\rightarrow\infty$, and with Ref. \cite{Ferraro:2007ec}, Equation (14) after turning off the external field $\bB'_0$.

The above are our results for the TE$_1$ wave polarized perpendicular to a constant transverse magnetic field in a parallel-plate waveguide in BLST electrodynamics. Eq. \eqref{om'Bx} shows that the cutoff frequency of TE$_1$ mode
\begin{equation}\label{omcBx}
\omega'_1\simeq k_{\bot}\left(1-\frac{A^2k_{\bot}^2}{4\tT}\right)
\end{equation}
is independent of the external field $\bB'_0$ but decreases with the amplitude $A$ of the TE$_1$ wave. The amplitude of the TE$_0$ mode is $\tT^{-1}Ak_{\|}B'_0/4$ times the amplitude of the TE$_1$ mode. As implied by Eqs. \eqref{vpvgBx} and \eqref{VBx}, the phase velocity decreases with the strength of external magnetic field and the amplitude of TE$_1$ wave, while the group velocity decreases with the strength of external magnetic field but increases with the amplitude of TE$_1$ wave. Combining Eqs. \eqref{vpvgBx} and \eqref{VBx}, we find
\begin{equation}\label{vgBx}
v_g^2\simeq1-\frac{k_{\bot}^2}{\omega'^2}+\frac{1}{\tT}\left[\frac{A^2k_{\bot}^4}{2\omega'^2}-B'^2_0\left(1-\frac{k_{\bot}^2}{\omega'^2}\right)\right].
\end{equation}

When deriving Eq. \eqref{ELBx}, we have taken $\tT$ as a constant implicitly. According to Eq. \eqref{sigma}, this is true at least if $\min F_{\mu\nu}F^{\mu\nu}\geq0$, or $B_0^2-(E_0+A\omega)^2\geq0$ for the $s_{11}$ mode. In the reference frame $S'$, it amounts to $B'_0/A\geq\omega'+\sqrt{\omega'^2-k_{\bot}^2}$ for the TE$_1$ mode, which can be satisfied by low-frequency waves. Once this condition is satisfied, we can set $\sigma=1$ and thus $\tT=e^{\gamma}T$.

\section{TE waves polarized parallel to a constant magnetic field background}\label{sect-By}
\subsection{Standing-wave solution}\label{subsect-Bystand}
In this section, we explore TE waves polarized parallel to a constant transverse magnetic field in a parallel-plate waveguide. Let us consider two perfectly conducting plates parallel to the $yz$-plane, located at $x=0$ and $x=L$ again. But this time the external electric and magnetic fields are parallel and perpendicular to the $x$-axis, respectively. Their directions are exchanged in comparison with the previous section. Since we are interested in an oscillating electric field along the external magnetic field, the simplest solution for the electromagnetic field in a convenient gauge \cite{Manojlovic:2020ndn} is of the form
\begin{eqnarray}\label{ABy}
\nonumber&&A_x(x,y,z,t)=-E_0t+B_0z,~~~~A_y(x,y,z,t)=u(x,t),\\
&&\Phi=A_z=0
\end{eqnarray}
with $u$ to be determined by field equations. We then have
\begin{eqnarray}
&&\bE=(E_0,-u_t,0),~~~~\bB=(0,B_0,u_x),\\
&&\frac{1}{4}F_{\mu\nu}F^{\mu\nu}=\frac{1}{2}\left(B_0^2-E_0^2+u_x^2-u_t^2\right),~~~~\frac{1}{4}F_{\mu\nu}\aF^{\mu\nu}=B_0u_t.
\end{eqnarray}
Substitution into Eq. \eqref{LBLST} shows that
\begin{equation}
\Lag=T-\sqrt{U}
\end{equation}
with the shorthand notation
\begin{equation}
U=T^2+T\cosh\gamma\left(B_0^2-E_0^2+u_x^2-u_t^2\right)-T\sinh\gamma\sqrt{\left(B_0^2-E_0^2+u_x^2-u_t^2\right)^2+4B_0^2u_t^2}-B_0^2u_t^2.
\end{equation}

Correspondingly, the Euler-Lagrange equation is
\begin{equation}\label{ELBy}
\frac{\partial\Lag}{\partial u}-\frac{\partial}{\partial x}\frac{\partial\Lag}{\partial u_x}-\frac{\partial}{\partial t}\frac{\partial\Lag}{\partial u_t}=0
\end{equation}
in which
\begin{eqnarray}
&&\frac{\partial\Lag}{\partial u_x}=-\frac{u_x}{\sqrt{U}}\left[T\cosh\gamma-\frac{T\sinh\gamma\left(B_0^2-E_0^2+u_x^2-u_t^2\right)}{\sqrt{\left(B_0^2-E_0^2+u_x^2-u_t^2\right)^2+4B_0^2u_t^2}}\right],\\
&&\frac{\partial\Lag}{\partial u_t}=\frac{u_t}{\sqrt{U}}\left[T\cosh\gamma-\frac{T\sinh\gamma\left(-B_0^2-E_0^2+u_x^2-u_t^2\right)}{\sqrt{\left(B_0^2-E_0^2+u_x^2-u_t^2\right)^2+4B_0^2u_t^2}}+B_0^2\right]
\end{eqnarray}
and $\partial\Lag/\partial u=0$. Multiplied by $U\sqrt{U}/T^3$, Eq. \eqref{ELBy} can be arranged in the form
\begin{equation}\label{ELC}
\frac{U\sqrt{U}}{T^3}\left(-\frac{\partial}{\partial x}\frac{\partial\Lag}{\partial u_x}-\frac{\partial}{\partial t}\frac{\partial\Lag}{\partial u_t}\right)=C_1u_{xx}-C_2u_{tt}+2C_3u_xu_tu_{xt}=0
\end{equation}
with
\begin{eqnarray}
\nonumber C_1&=&\left(\cosh\gamma-\sinh\gamma\cos\theta-\frac{2u_x^2\sinh\gamma\sin^2\theta}{\sqrt{\Xi}}\right)\left(1-\frac{B_0^2u_t^2}{T^2}\right)+\frac{\left(B_0^2-E_0^2-u_t^2\right)(\cosh\gamma-\sinh\gamma\cos\theta)^2}{T}\\
\nonumber&&+\frac{2B_0u_t\sinh\gamma\sin\theta(\cosh\gamma-\sinh\gamma\cos\theta)}{T}-\frac{2u_x^2\sinh\gamma\sin^2\theta(\cosh\gamma\cos\theta-\sinh\gamma)}{T}\\
&=&1-\gamma\cos\theta-\frac{2\gamma u_x^2\sin^2\theta}{\sqrt{\Xi}}+\frac{B_0^2-E_0^2-u_t^2}{T}+\mathcal{O}\left(\gamma^2,T^{-2},\gamma T^{-1}\right),
\end{eqnarray}
\begin{eqnarray}
\nonumber C_2&=&\left(\cosh\gamma-\sinh\gamma\cos\theta\right)\left[1+\frac{B_0^2\left(B_0^2-E_0^2+u_x^2\right)}{T^2}\right]+\frac{\left(2B_0^2-E_0^2+u_x^2\right)(\cosh\gamma-\sinh\gamma\cos\theta)^2}{T}\\
\nonumber&&+\frac{4B_0^2\sinh\gamma\cos\theta(\cosh\gamma-\sinh\gamma\cos\theta)}{T}+\frac{2\left(u_x^2-E_0^2\right)\sinh\gamma\sin^2\theta(\cosh\gamma\cos\theta-\sinh\gamma)}{T}\\
\nonumber&&-\frac{2B_0^2\sinh^2\gamma\sin^2\theta}{T}+\frac{2\sinh\gamma\left[B_0^2+\left(u_x^2-E_0^2\right)\sin^2\theta\right]}{\sqrt{\Xi}}\left(1-\frac{B_0^2u_t^2}{T^2}\right)\\
&=&1-\gamma\cos\theta+\frac{2B_0^2-E_0^2+u_x^2}{T}+\frac{2\gamma\left[B_0^2+\left(u_x^2-E_0^2\right)\sin^2\theta\right]}{\sqrt{\Xi}}+\mathcal{O}\left(\gamma^2,T^{-2},\gamma T^{-1}\right),
\end{eqnarray}
\begin{eqnarray}
\nonumber C_3&=&\frac{B_0^2(\cosh\gamma-\sinh\gamma\cos\theta)}{T^2}+\frac{\cosh^2\gamma-2\sinh\gamma\cosh\gamma\cos^3\theta+\sinh^2\gamma\left(3\cos^2\theta-2\right)}{T}\\
\nonumber&&+\frac{2B_0^2\sinh\gamma\left[\cosh\gamma\left(2\cos^2\theta+1\right)-3\sinh\gamma\cos\theta\right]}{T\sqrt{\Xi}}+2\sinh\gamma\left(\frac{2B_0^2\cos\theta}{\Xi}+\frac{\sin^2\theta}{\sqrt{\Xi}}\right)\left(1-\frac{B_0^2u_t^2}{T^2}\right)\\
&=&\frac{1}{T}+2\gamma\left(\frac{2B_0^2\cos\theta}{\Xi}+\frac{\sin^2\theta}{\sqrt{\Xi}}\right)+\mathcal{O}\left(\gamma^2,T^{-2},\gamma T^{-1}\right)
\end{eqnarray}
and
\begin{eqnarray}
\cos\theta&=&\frac{B_0^2-E_0^2+u_x^2-u_t^2}{\sqrt{\Xi}},~~~~\sin\theta=\frac{-2B_0u_t}{\sqrt{\Xi}},\\
\Xi&=&\left(B_0^2-E_0^2+u_x^2-u_t^2\right)^2+4B_0^2u_t^2.
\end{eqnarray}

The field equation \eqref{ELC} is too complicated to solve, even approximately with the Poincar\'{e}-Lindstedt method \cite{Manojlovic:2020ndn}. However, it is not hard to see that the boundary conditions $u(0,t)=u(L,t)=0$ are compatible with Eq. \eqref{ELC} by decomposing its standing-wave solution $u(x,t)~(0\leq x\leq L)$ into series of $s_{nm}(x,t)$, $c_{nm}(x,t)~(n\geq1)$ defined by Eqs. \eqref{snm}, \eqref{cnm}. Formally the solution of Eq. \eqref{ELC} can be arranged order by order as
\begin{eqnarray}
\label{uBy}u&=&u^{(0,0)}+u^{(1,0)}+u^{(0,1)}+u^{(2,0)}+u^{(1,1)}+u^{(0,2)}+\cdots,\\
\omega^2&=&\omega^2_{(0,0)}+\omega^2_{(1,0)}+\omega^2_{(0,1)}+\omega^2_{(2,0)}+\omega^2_{(1,1)}+\omega^2_{(0,2)}+\cdots,
\end{eqnarray}
where $\omega^2_{(i,j)}~(i,j\geq0)$ is of order $\gamma^iT^{-j}$. Under the boundary conditions $u(0,t)=u(L,t)=0$, each of $u^{(i,j)}(x,t)~(i,j\geq0)$ is a linear combination of functions $s_{nm}(x,t),~c_{nm}(x,t)~(n\geq1)$ with coefficients of order $\gamma^iT^{-j}$, and it is reasonable to assume that the $s_{11}(x,t)$ mode does not appear in $u^{(i,j)}(x,t)~(i+j\geq1)$. We will not work out $u^{(i,j)}(x,t)~(i+j\geq1)$, because they are more complicated but less interesting than the dispersion relation of the TE$_1$ mode. The $\mathcal{O}(\gamma,1/T)$ dispersion relation can be derived by inserting the seed solution
\begin{equation}\label{uom00By}
u^{(0,0)}(x,t)=A\sin(kx)\cos(\omega t)=\frac{A}{2}s_{11}(x,t),~~~~\omega^2_{(0,0)}=k^2
\end{equation}
into Eq. \eqref{ELC} multiplied by $\Xi\sqrt{\Xi}$ and then extracting the coefficient of $s_{11}$,
\begin{equation}\label{coeBy}
\omega^2-k^2-\omega^2_{(1,0)}+\frac{k^2}{T}\left(B_0^2+\frac{A^2k^2}{2}\right)+\mathcal{O}\left(\gamma^2,T^{-2},\gamma T^{-1}\right),
\end{equation}
in which
\begin{eqnarray}
\label{om10By}\omega^2_{(1,0)}&=&-\frac{\gamma L^2B_0^2k^2\left[3A^4k^4+8A^2k^2\left(B_0^2-E_0^2\right)+8\left(B_0^2-E_0^2\right)^2\right]}{\displaystyle{4\int_{-L}^{L}\int_{-L}^{L}\sin^2(kx)\cos^2(kt)\Xi_{(0,0)}^{(0,0)3/2}dxdt}},\\
\label{Xi0000}\Xi_{(0,0)}^{(0,0)}&=&\left[A^2k^2\cos^2(kt)-A^2k^2\sin^2(kx)+B_0^2-E_0^2\right]^2+4A^2k^2B_0^2\sin^2(kx)\sin^2(kt).
\end{eqnarray}
Note that were there $s_{11}(x,t)$ mode in $u^{(i,j)}(x,t)~(i+j\geq1)$, it can be absorbed into $u^{(0,0)}(x,t)$ by adjusting the amplitude $A$ in Eq. \eqref{uom00By}.

Strictly speaking, we ought to put $u=u^{(0,0)}+u^{(1,0)}+u^{(0,1)}$, $\omega^2=\omega^2_{(0,0)}+\omega^2_{(1,0)}+\omega^2_{(0,1)}$ into Eq. \eqref{ELC} to extract the coefficient of $s_{11}$. But $u^{(1,0)}$ and $u^{(0,1)}$ do not contain $s_{11}$ by assumption, so they have no influence on $\mathcal{O}(\gamma,1/T)$ terms in Eq. \eqref{coeBy}. Particularly, we have
\begin{equation}
\Xi\sqrt{\Xi}\left(u_{xx}-u_{tt}\right)=\Xi^{(0,0)3/2}_{(0,0)}\left(\omega^2-k^2\right)u^{(0,0)}+\mathcal{O}\left(\gamma^2,T^{-2},\gamma T^{-1}\right)
\end{equation}
 as long as $\omega^2_{(0,0)}=k^2$. To satisfy Eq. \eqref{ELC}, the coefficient of $s_{11}$ should be equal to zero. As a result, the dispersion relation of the $s_{11}$ mode of standing wave is
\begin{equation}\label{omBy}
\omega^2=k^2+\omega^2_{(1,0)}-\frac{k^2}{T}\left(B_0^2+\frac{A^2k^2}{2}\right)+\mathcal{O}\left(\gamma^2,T^{-2},\gamma T^{-1}\right).
\end{equation}

\subsection{Propagating TE$_1$ mode}\label{subsect-ByTE1}
Parallel to Sec. \ref{subsect-BxTE1}, we can do a Lorentz boost through the coordinate transformation \eqref{boost} to turn the standing-wave solution \eqref{uBy} into propagating waves. Then the phase of the temporal factor changes to $\omega't'-k_{\|}z'$ but the phase of the transversal sector remains invariant. The frequency and the wave vector are given by Eq. \eqref{4wav}.

Under the Lorentz boost \eqref{boost}, the electromagnetic field \eqref{ABy} is transformed into
\begin{eqnarray}
\nonumber&&A_{x'}=-\frac{E_0+VB_0}{\sqrt{1-V^2}}t'+\frac{B_0+VE_0}{\sqrt{1-V^2}}z',\\
&&A_{y'}=u\left(x',\frac{t'-Vz'}{\sqrt{1-V^2}}\right),~~~~\Phi=A_{z'}=0.
\end{eqnarray}
In this section, we are interested in TE waves polarized parallel to a constant external magnetic field. For this reason, we set $E_0=-VB_0$ \cite{Jackson:1999ny} to get rid of an external electric field in the new reference frame. Then we have
\begin{equation}
\bE'=(0,-u_{t'},0),~~~~\bB'=\left(-u_{z'},B'_0,u_{x'}\right),
\end{equation}
where the external magnetic field is $B'_0=B_0\sqrt{1-V^2}$ in the $y'$-direction as expected. We have illustrated the experimental configuration in the lower panel of Fig. \ref{fig-conf}.

Again the $s_{11}$ mode of standing wave is turned into the TE$_1$ mode of propagating wave, whose dispersion relation can be gained by reforming Eq. \eqref{omBy} as
\begin{equation}\label{om'By}
\omega'^2-k_{\|}^2=k_{\bot}^2+\omega^2_{(1,0)}-\frac{1}{T}\left(\omega'^2B'^2_0+\frac{A^2k_{\bot}^4}{2}\right)+\mathcal{O}\left(\gamma^2,T^{-2},\gamma T^{-1}\right),
\end{equation}
with the wave four-vector defined in Eq. \eqref{4wav}. From this dispersion relation, one can get the phase velocity and group velocity of the TE$_1$ mode
\begin{equation}\label{vpvgBy}
v_p=\frac{1}{V},~~~~v_g=\left(1-\frac{d\omega^2_{(1,0)}}{d\left(\omega'^2\right)}+\frac{B'^2_0}{\tT}\right)^{-1}V+\mathcal{O}\left(\gamma^2,T^{-2},\gamma T^{-1}\right),
\end{equation}
respectively, where
\begin{equation}\label{VBy}
V^2=1-\frac{k_{\bot}^2}{\omega'^2}-\frac{\omega^2_{(1,0)}}{\omega'^2}+\frac{1}{T}\left(B'^2_0+\frac{A^2k_{\bot}^4}{2\omega'^2}\right)+\mathcal{O}\left(\gamma^2,T^{-2},\gamma T^{-1}\right).
\end{equation}
In the above, $\omega^2_{(1,0)}$ should be understood as a function of $B'_0$, $\omega'$ and $k_{\bot}$. It can be derived from Eq. \eqref{om10By} as
\begin{equation}\label{om10'By}
\omega^2_{(1,0)}=-\frac{\gamma\pi^2\omega'^2b^2\left(3+8b^2+8b^4\right)}{\displaystyle{4\int_{-\pi}^{\pi}\int_{-\pi}^{\pi}\sin^2\xi\cos^2\tau\left[\left(\cos^2\tau-\sin^2\xi+b^2\right)^2+4\omega'^2k_{\bot}^{-2}b^2\sin^2\xi\sin^2\tau\right]^{3/2}d\xi d\tau}}.
\end{equation}
with the notation $b=B'_0/(Ak_{\bot})$. Setting $\gamma=0$, Eq. \eqref{VBy} is consistent with Ref. \cite{Sorokin:2021tge}, Equation (54) in the limit $L\rightarrow\infty$, and with Ref. \cite{Ferraro:2007ec}, Equation (14) after turning off the external field $\bB'_0$. Because $\omega^2_{(1,0)}$ is the $\mathcal{O}(\gamma)$ term in the series expansion of $\omega^2$, it need not be nonnegative. In fact, Eq. \eqref{om10'By} indicates that $\omega^2_{(1,0)}$ is positive as long as $b\neq0$ and $\gamma>0$.

\begin{figure}
\centering
\includegraphics[width=0.45\textwidth]{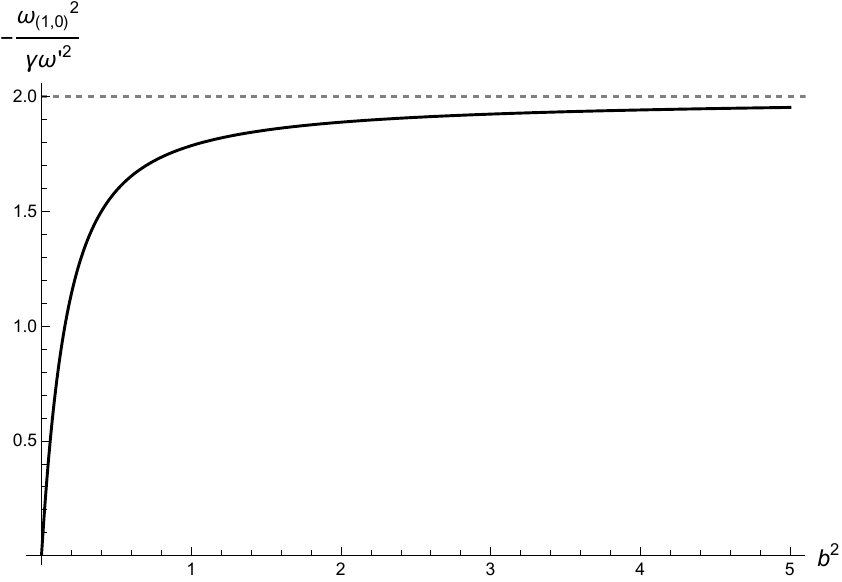}
\caption{Illustration of $-\omega^2_{(1,0)}/(\gamma\omega'^2)$ in the region $0\leq b^2\leq5$ after fixing $\omega'=k_{\bot}$, as depicted by the solid curve according to Eq. \eqref{om10'By}. In agreement with Eq. \eqref{om10'By}, the curve tends to $2$ in the limit $b\rightarrow\infty$, and to $0$ at $b=0$. The dotted asymptotic line corresponds to $-\omega^2_{(1,0)}/(\gamma\omega'^2)=2$.}\label{fig-2Dom10}
\end{figure}

\begin{figure}
\centering
\includegraphics[width=0.45\textwidth]{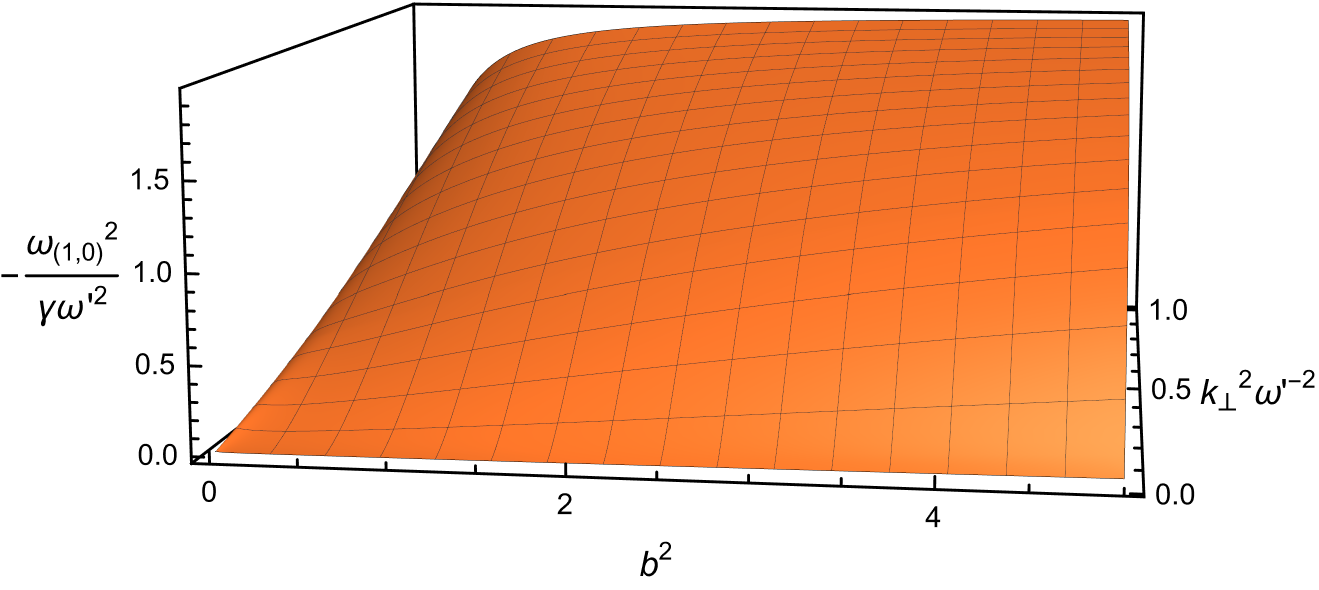}
\caption{Illustration of $-\omega^2_{(1,0)}/(\gamma\omega'^2)$ in the contour $0<k_{\bot}^2\omega'^{-2}\leq1$, $0\leq b^2\leq5$ according to Eq. \eqref{om10'By}. In agreement with Eq. \eqref{om10'By}, it tends to $0$ in the limit $k_{\bot}/\omega'=0$ or $b=0$ with the other parameter fixed.}\label{fig-3Dom10}
\end{figure}

\begin{figure}
\centering
\includegraphics[width=0.45\textwidth]{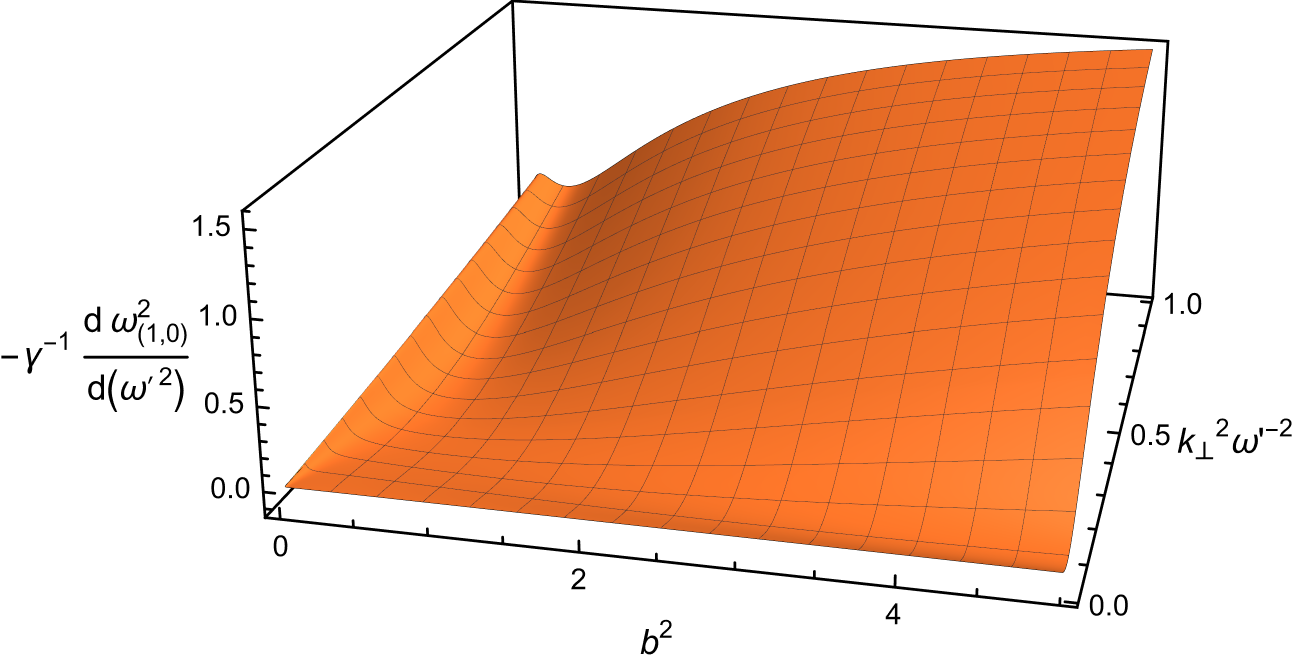}\\
\includegraphics[width=0.45\textwidth]{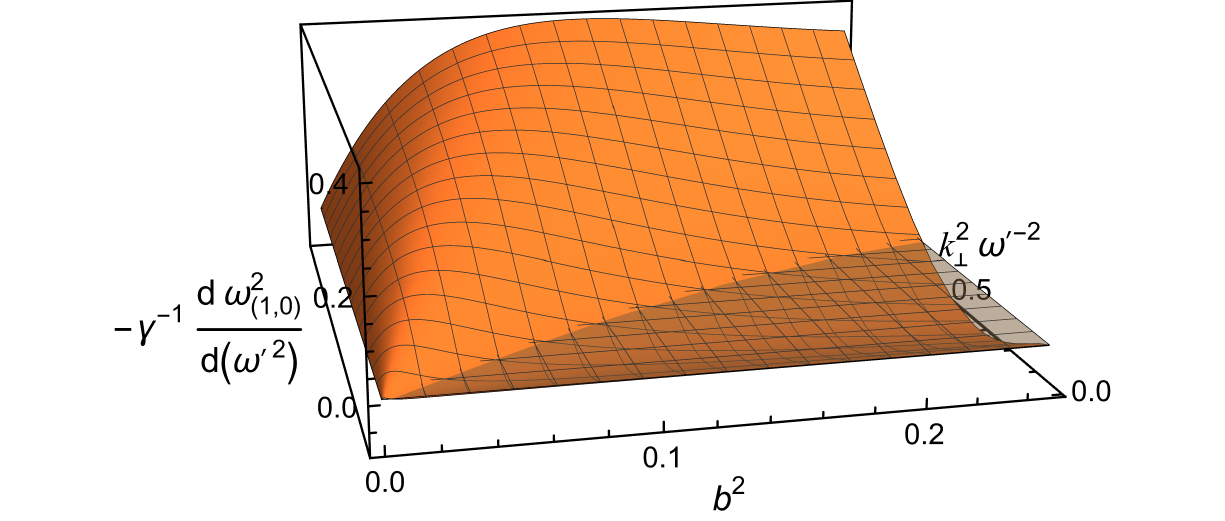}
\caption{Illustration of $-\gamma^{-1}d\omega^2_{(1,0)}/d(\omega'^2)$ in the contour $0<k_{\bot}^2\omega'^{-2}\leq1$, $0\leq b^2\leq5$ (upper panel) or $0\leq b^2\leq0.25$ (lower panel). It tends to $0$ in the limit $k_{\bot}/\omega'=0$ or $b=0$ with the other parameter fixed. The shaded plane in the lower subfigure represents $-\gamma^{-1}d\omega^2_{(1,0)}/d(\omega'^2)=0$.}\label{fig-dom10}
\end{figure}

\begin{figure}
\centering
\includegraphics[width=0.45\textwidth]{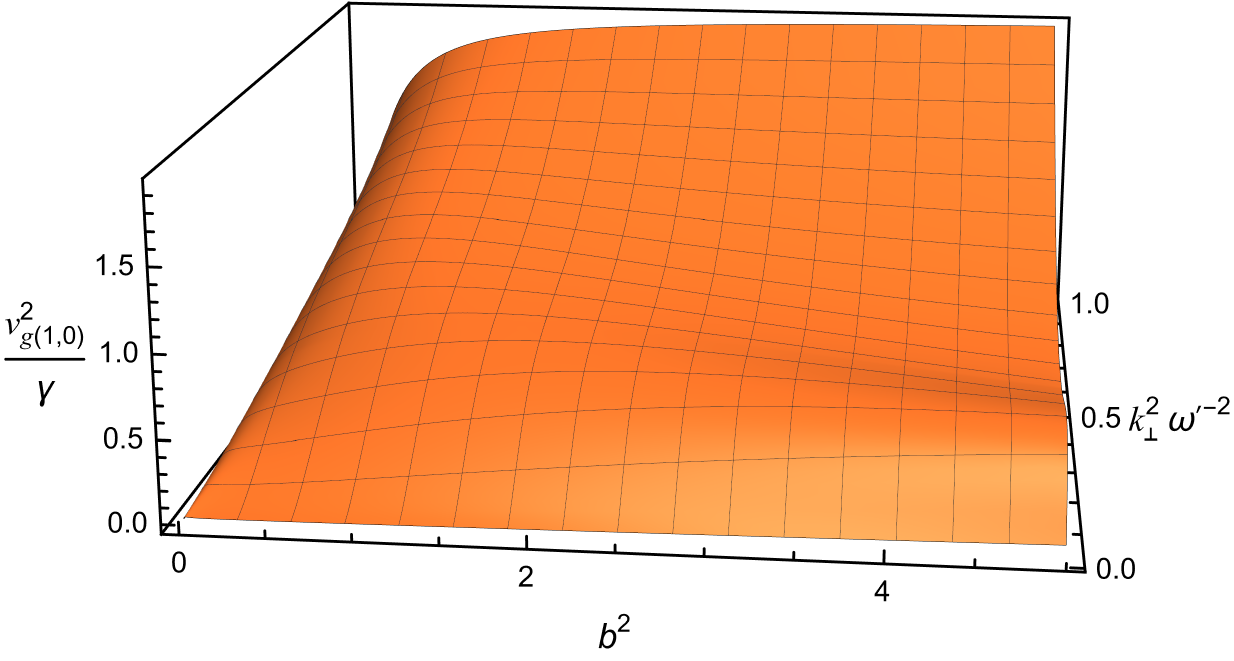}
\caption{Illustration of $v^2_{g(1,0)}/\gamma$ in the contour $0<k_{\bot}^2\omega'^{-2}\leq1$, $0\leq b^2\leq5$ according to Eq. \eqref{vg10By}. It tends to $0$ in the limit $k_{\bot}/\omega'=0$ or $b=0$ with the other parameter fixed. In the limit $k_{\bot}/\omega'=1$, it reduces to the curve in Fig. \ref{fig-2Dom10}}\label{fig-vg10}
\end{figure}

The above are our results for the TE$_1$ wave polarized parallel to a constant transverse magnetic field in a parallel-plate waveguide in BLST electrodynamics. Neglecting $\mathcal{O}(\gamma^2,T^{-2},\gamma T^{-1})$ terms, Eq. \eqref{om'By} means that the cutoff frequency of TE$_1$ mode is
\begin{equation}\label{omcBy}
\omega'_1\simeq k_{\bot}\left(1+\left.\frac{\omega^2_{(1,0)}}{2\omega'^2}\right|_{\omega'=k_{\bot}}-\frac{2B'^2_0+A^2k_{\bot}^2}{4T}\right).
\end{equation}
Influenced by the $\mathcal{O}(T^{-1})$ correction, the cutoff frequency decreases with the strength of external magnetic field $B'_0$ and the amplitude $A$ of TE$_1$ wave. Fixing $\omega'=k_{\bot}$, we plot $-\omega^2_{(1,0)}/(\gamma\omega'^2)$ in the region $0\leq b^2\leq5$ in Fig. \ref{fig-2Dom10}, which shows that the cutoff frequency decreases with the ratio of external field strength to wave amplitude $B'_0/A$ as a result of the $\mathcal{O}(\gamma)$ correction. In Fig. \ref{fig-3Dom10}, we depict $-\omega^2_{(1,0)}/(\gamma\omega'^2)$ in the contour $0<k_{\bot}^2\omega'^{-2}\leq1$, $0\leq b^2\leq5$. The $\mathcal{O}(T^{-1})$ correction in Eq. \eqref{VBy} suggests that the phase velocity decreases with the strength of external magnetic field and the amplitude of TE$_1$ wave, while the $\mathcal{O}(\gamma)$ correction indicates that the phase velocity decreases with the ratio $B'_0/A$. Making use of Eqs. \eqref{vpvgBy}, \eqref{VBy}, we can obtain
\begin{eqnarray}
\nonumber v_g^2&\simeq&1-\frac{k_{\bot}^2}{\omega'^2}+2\left(1-\frac{k_{\bot}^2}{\omega'^2}\right)\frac{d\omega^2_{(1,0)}}{d\left(\omega'^2\right)}-\frac{\omega^2_{(1,0)}}{\omega'^2}\\
\label{vgBy}&&+\frac{1}{T}\left[\frac{A^2k_{\bot}^4}{2\omega'^2}-B'^2_0\left(1-\frac{2k_{\bot}^2}{\omega'^2}\right)\right].
\end{eqnarray}
The $\mathcal{O}(T^{-1})$ correction in this equation suggests that the group velocity increases with the amplitude of TE$_1$ wave, but its dependence on the strength of external magnetic field is sensitive to the value of $1-2k_{\bot}^2/\omega'^2$.  In Fig. \ref{fig-dom10}, we illustrate $-\gamma^{-1}d\omega^2_{(1,0)}/d(\omega'^2)$ in the contour $0<k_{\bot}^2\omega'^{-2}\leq1$, $0\leq b^2\leq5$. Similar plots are displayed in Fig. \ref{fig-vg10} for the $\mathcal{O}\left(\gamma\right)$ correction in $v_g^2$,
\begin{equation}\label{vg10By}
v^2_{g(1,0)}=2\left(1-\frac{k_{\bot}^2}{\omega'^2}\right)\frac{d\omega^2_{(1,0)}}{d\left(\omega'^2\right)}-\frac{\omega^2_{(1,0)}}{\omega'^2}.
\end{equation}
This correction does not have a monotonic dependence on the ratio $B'_0/A$. Near the cutoff frequency, the $\mathcal{O}\left(\gamma\right)$ correction is approximately given by the last term in Eq. \eqref{vg10By}.

\section{Testing BLST electrodynamics simply with $\omega'=k_{\bot}$ TE$_1$ wave}\label{sect-test}
In the BLST electrodynamics, we have studied the properties of TE$_1$ wave described by the term $A\sin(k_{\bot}x')\cos(\omega't'-k_{\|}z')$ with $k_{\bot}=\pi/L$ in the $y'$-component of the electromagnetic potential. Discussions on the general implication of our results will be deferred to Sec. \ref{sect-con}. In the current section, as a simple guide to the waveguide test of BLST theory, we focus on a specific TE$_1$ wave of frequency $\omega'=k_{\bot}$. This is exactly the cutoff frequency of TE$_1$ mode in Maxwell electrodynamics. However, it is higher than the cutoff frequency of TE$_1$ wave in BLST electrodynamics, as suggested by Eqs. \eqref{omcBx} and \eqref{omcBy}. In other words, in the presence of an external magnetic field, the $\omega'=k_{\bot}$ TE$_1$ wave can propagate in the waveguide unless both $\gamma$ and $T^{-1}$ vanish. Remarkably, in the waveguide experiment, constraints on $T$ and $\gamma$ can be extracted simply from the group velocity of the TE$_1$ wave of frequency $\omega'=k_{\bot}$.

The experimental configurations have been displayed in Fig. \ref{fig-conf}. In the first configuration, the external magnetic field is normal to the waveguide plates. In practice, this can be built by attaching opposite magnetic poles to the parallel plates, which are made of a paramagnetic or ferromagnetic good conductor of electricity. For the TE$_1$ wave of frequency $\omega'=k_{\bot}$, Eq. \eqref{vgBx} reduces to
\begin{equation}\label{vgcBx}
\left.v_g^2\right|_{\omega'=k_{\bot}}\simeq\frac{A^2k_{\bot}^2}{2T},
\end{equation}
where $\simeq$ stands for equivalence up to $\mathcal{O}(\gamma^2,T^{-2},\gamma T^{-1})$ terms. In the second configuration, the external field is parallel to the plates. This configuration is more accessible to a powerful magnetic field. Setting $\omega'=k_{\bot}$, one can simplify Eq. \eqref{vgBy} significantly,
\begin{equation}\label{vgcBy}
\left.v_g^2\right|_{\omega'=k_{\bot}}\simeq-\left.\frac{\omega^2_{(1,0)}}{\omega'^2}\right|_{\omega'=k_{\bot}}+\frac{A^2k_{\bot}^2}{2T}+\frac{B'^2_0}{T}.
\end{equation}

The group velocity $v_g$ in Eq. \eqref{vgcBx} is to be measured in the first configuration, whereas $v_g$ in Eq. \eqref{vgcBy} is to be measured in the second configuration. After the measurements of $v_g$, the constraint on $T$ can be extracted from Eq. \eqref{vgcBx}, and the constraint on $\gamma$ can be derived by combining Eqs. \eqref{vgcBx} and \eqref{vgcBy}.

Ref. \cite{Ellis:2017edi} put a lower limit $T^{1/4}\gtrsim100~\mathrm{GeV}$ on Born-Infeld theory with the ATLAS measurement of light-by-light scattering in Pb-Pb collisions at the LHC \cite{ATLAS:2017fur}. Unfortunately, because of the kinematic cuts made in the ATLAS analysis, their limit does not apply to a range of values of  $T^{1/4}\lesssim10~\mathrm{GeV}$. In Ref. \cite{NiauAkmansoy:2017kbw}, a limit $T^{1/2}>1.07\times10^{21}~\mathrm{V/m}$ was obtained, six orders of magnitude weaker than that of Ref. \cite{Ellis:2017edi}.

Taking $Ak_{\bot}\simeq10~\mathrm{teslas}$ and $T^{1/2}>1.07\times10^{21}~\mathrm{V/m}$, we find accordingly $A^2k_{\bot}^2/T\lesssim10^{-23}$, or equivalently $\left.v_g\right|_{\omega'=k_{\bot}}\lesssim10^{-3}~\mathrm{m/s}$ for Eq. \eqref{vgcBx}. On the right hand side of Eq. \eqref{vgcBy}, the third term $B'^2_0/T$ can be estimated in the same way. But the first term is different. It is proportional to the ModMax constant $\gamma$, which is largely unconstrained in the literature \cite{Hoshimov:2025tdx}, and the coefficient of proportionality is fully determined by the ratio $b=B'_0/(Ak_{\bot})$. As depicted by Fig. \ref{fig-2Dom10}, the value of this term tends to $2\gamma$ for $b\gtrsim1$, but is of order $b^2\gamma$ if $b\ll1$. These estimates must be taken into consideration in an elaborate experiment.


\section{Conclusion}\label{sect-con}
The BLST electrodynamics is a unified framework to study Maxwell, Born-Infeld, Bialynicki-Birula and ModMax theories. Its rich implications should be investigated not only theoretically but also experimentally. In this framework, after imposing a constant external magnetic field normal to the propagation direction, we have explored the TE$_1$ waves in a parallel-plate waveguide, including their phase velocity $v_p$, group velocity $v_g$ and cutoff frequency $\omega'_1$. The three physical quantities all receive corrections from terms characterized by Born-Infeld constant $T$ and ModMax constant $\gamma$. Hunting for such corrections in experiments would help us to test BLST electrodynamics.

In Sec. \ref{sect-Bx}, the external magnetic field is perpendicular to the direction of the electric polarization of guided waves. Under the condition of $B'_0/A\geq\omega'+\sqrt{\omega'^2-k_{\bot}^2}$, we found the BLST corrections to $v_p$, $v_g$ and $\omega'_1$ of TE$_1$ waves have the same forms as corrections in Born-Infeld electrodynamics, except for that $T$ is replaced by  $\tT=e^{\gamma}T$. Consequently, in this case, the waveguide experiment cannot distinguish between BLST theory and Born-Infeld theory, but it can put a bound on the value of $e^{\gamma}T$, where the constraints on $\gamma$ and $T$ are correlated.

In Sec. \ref{sect-By}, the external field is along the electric polarization direction of guided waves. This configuration has richer phenomena. We found the Born-Infeld corrections to $v_p$, $v_g$ and $\omega'_1$ of TE$_1$ waves are different from the ModMax corrections. The ModMax corrections depend simply on the ratio of external magnetic field strength $B'_0$ to TE wave amplitude $A$, while the Born-Infeld corrections are linear combinations of $B'^2_0$ and $A^2$. Specifically, if the external magnetic field is turned off, then the $\mathcal{O}(T^{-1})$ corrections will vanish but the $\mathcal{O}(\gamma)$ corrections will survive. Therefore, in this configuration, the experiment can disentangle the constraints on $\gamma$ and $T$, suited to test the BLST electrodynamics.

Although we have restricted our investigations to parallel-plane waveguides, the results can be also applied to rectangular waveguides, in which the TE$_{1,0}$ and the TE$_{0,1}$ modes are identical to the TE$_1$ mode in the two configurations in this article.



\section*{Acknowledgments}
This work is sponsored by the Natural Science Foundation of Shanghai (Grant No. 24ZR1419300) and the National Natural Science Foundation of China (Grant No. 12575060).


\section*{Data availability}
No data was used for the research described in the article.


\begin{thebibliography}{99}


\bibitem{Born:1934gh}
M.~Born and L.~Infeld,
Proc. Roy. Soc. Lond. A \textbf{144}, no.852, 425-451 (1934)

\bibitem{Heisenberg:1936nmg}
W.~Heisenberg and H.~Euler,
Z. Phys. \textbf{98}, no.11-12, 714-732 (1936)
[arXiv:physics/0605038 [physics]].

\bibitem{BialynickiBirula:1984tx}
I.~Bialynicki-Birula,
``Nonlinear Electrodynamics: Variations on a theme by Born and Infeld'',
in {\sl Quantum Theory Of Particles and Fields: birthday volume dedicated to  Jan \L{}opusza\'nski}, eds B. Jancewicz and J.  Lukierski, pp 31-48, (World Scientific, 1983).

\bibitem{Bialynicki-Birula:1992rcm}
I.~Bialynicki-Birula,
Acta Phys. Polon. B \textbf{23}, 553-559 (1992)

\bibitem{Soleng:1995kn}
H.~H.~Soleng,
Phys. Rev. D \textbf{52}, 6178-6181 (1995)
[arXiv:hep-th/9509033 [hep-th]].

\bibitem{Jing:2011vz}
J.~Jing, Q.~Pan and S.~Chen,
JHEP \textbf{11}, 045 (2011)
[arXiv:1106.5181 [hep-th]].

\bibitem{Gaete:2013dta}
P.~Gaete and J.~Helay{\"e}l-Neto,
Eur. Phys. J. C \textbf{74}, no.3, 2816 (2014)
[arXiv:1312.5157 [hep-th]].

\bibitem{Kruglov:2014hpa}
S.~I.~Kruglov,
Annals Phys. \textbf{353}, 299-306 (2014)
[arXiv:1410.0351 [physics.gen-ph]].

\bibitem{Bandos:2020jsw}
I.~Bandos, K.~Lechner, D.~Sorokin and P.~K.~Townsend,
Phys. Rev. D \textbf{102}, 121703 (2020)
[arXiv:2007.09092 [hep-th]].

\bibitem{Kosyakov:2020wxv}
B.~P.~Kosyakov,
Phys. Lett. B \textbf{810}, 135840 (2020)
[arXiv:2007.13878 [hep-th]].

\bibitem{Gullu:2020ant}
I.~Gullu and S.~H.~Mazharimousavi,
Phys. Scripta \textbf{96}, no.4, 045217 (2021)
[arXiv:2009.08665 [gr-qc]].

\bibitem{Russo:2025fuc}
J.~G.~Russo and P.~K.~Townsend,
JHEP \textbf{10}, 120 (2025)
[arXiv:2505.08869 [hep-th]].

\bibitem{Dunne:2012vv}
G.~V.~Dunne,
Int. J. Mod. Phys. A \textbf{27}, 1260004 (2012)
[arXiv:1202.1557 [hep-th]].

\bibitem{Sorokin:2021tge}
D.~P.~Sorokin,
Fortsch. Phys. \textbf{70}, no.7-8, 2200092 (2022)
[arXiv:2112.12118 [hep-th]].

\bibitem{Jackson:1999ny}
J.~D.~Jackson,
``Classical Electrodynamics,'' Wiley, New York, 1999.

\bibitem{Russo:2022qvz}
J.~G.~Russo and P.~K.~Townsend,
JHEP \textbf{01}, 039 (2023)
[arXiv:2211.10689 [hep-th]].

\bibitem{Mezincescu:2023zny}
L.~Mezincescu, J.~G.~Russo and P.~K.~Townsend,
JHEP \textbf{02}, 186 (2024)
[arXiv:2311.04278 [hep-th]].

\bibitem{Bialynicka-Birula:1970nlh}
Z.~Bialynicka-Birula and I.~Bialynicki-Birula,
Phys. Rev. D \textbf{2}, 2341-2345 (1970)

\bibitem{Adler:1971wn}
S.~L.~Adler,
Annals Phys. \textbf{67}, 599-647 (1971)

\bibitem{Mendonca:2006pg}
J.~T.~Mendonca, J.~Dias de Deus and P.~Castelo Ferreira,
Phys. Rev. Lett. \textbf{97}, 100403 (2006)
[erratum: Phys. Rev. Lett. \textbf{97}, 269901 (2006)]
[arXiv:hep-ph/0606099 [hep-ph]].

\bibitem{Adler:2006zs}
S.~L.~Adler,
J. Phys. A \textbf{40}, F143-F152 (2007)
[arXiv:hep-ph/0611267 [hep-ph]].

\bibitem{Aleksandrov:2021jtj}
I.~A.~Aleksandrov and V.~M.~Shabaev,
Opt. Spectrosc. \textbf{129}, no.8, 890-895 (2021)


\bibitem{Denisov:2016pfu}
V.~I.~Denisov, E.~E.~Dolgaya and V.~A.~Sokolov,
JHEP \textbf{05}, 105 (2017)
[arXiv:1612.09086 [hep-ph]].

\bibitem{Bragin:2017yau}
S.~Bragin, S.~Meuren, C.~H.~Keitel and A.~Di Piazza,
Phys. Rev. Lett. \textbf{119}, no.25, 250403 (2017)
[arXiv:1704.05234 [hep-ph]].


\bibitem{Novello:1999pg}
M.~Novello, V.~A.~De Lorenci, J.~M.~Salim and R.~Klippert,
Phys. Rev. D \textbf{61}, 045001 (2000)
[arXiv:gr-qc/9911085 [gr-qc]].

\bibitem{Aiello:2006cz}
M.~Aiello, G.~Bengochea and R.~Ferraro,
Phys. Lett. A \textbf{361}, 9-12 (2007)
[arXiv:hep-th/0607072 [hep-th]].

\bibitem{Dinu:2013gaa}
V.~Dinu, T.~Heinzl, A.~Ilderton, M.~Marklund and G.~Torgrimsson,
Phys. Rev. D \textbf{89}, no.12, 125003 (2014)
[arXiv:1312.6419 [hep-ph]].

\bibitem{Guzman-Herrera:2020ffc}
E.~Guzman-Herrera and N.~Breton,
Eur. Phys. J. C \textbf{81}, no.2, 115 (2021)
[arXiv:2008.10739 [hep-th]].

\bibitem{Neves:2021tbt}
M.~J.~Neves, J.~B.~de Oliveira, L.~P.~R.~Ospedal and J.~A.~Helay{\"e}l-Neto,
Phys. Rev. D \textbf{104}, no.1, 015006 (2021)
[arXiv:2101.03642 [hep-th]].


\bibitem{Perez-Garcia:2022kvz}
M.~{\'A}.~P{\'e}rez-Garc{\'\i}a, A.~P{\'e}rez Mart{\'\i}nez and E.~Rodr{\'\i}guez Querts,
Eur. Phys. J. C \textbf{83}, no.8, 746 (2023)
[arXiv:2212.05086 [hep-ph]].


\bibitem{Santos:2023ooc}
G.~R.~Santos and M.~J.~Neves,
Int. J. Mod. Phys. A \textbf{39}, no.07n08, 2450045 (2024)
[arXiv:2311.02080 [physics.class-ph]].


\bibitem{Neves:2022jqq}
M.~J.~Neves, P.~Gaete, L.~P.~R.~Ospedal and J.~A.~Helay{\"e}l-Neto,
Phys. Rev. D \textbf{107}, no.7, 075019 (2023)
[arXiv:2209.09361 [hep-th]].

\bibitem{Cruz:2025xut}
T.~W.~Cruz, V.~A.~De Lorenci, E.~Guzm{\'a}n-Herrera and C.~C.~H.~Ribeiro,
Eur. Phys. J. C \textbf{85}, no.7, 761 (2025)
[arXiv:2503.24369 [physics.gen-ph]].

\bibitem{Shi:2024nmx}
Y.~Shi and T.~Wang,
Nucl. Phys. B \textbf{1020}, 117166 (2025)
[arXiv:2403.20044 [hep-ph]].

\bibitem{Shi:2025bri}
Y.~Shi and T.~Wang,
J. East China Normal Univer. Natur. Sci. \textbf{107}, no.3, 0013 (2025)

\bibitem{Fouche:2016lne}
M.~Fouche, R.~Battesti and C.~Rizzo,
Phys. Rev. D \textbf{93}, 093020 (2016)
[erratum: Phys. Rev. D \textbf{95}, 099902 (2017)]
[arXiv:1605.04102 [physics.optics]].

\bibitem{Ellis:2017edi}
J.~Ellis, N.~E.~Mavromatos and T.~You,
Phys. Rev. Lett. \textbf{118}, no.26, 261802 (2017)
[arXiv:1703.08450 [hep-ph]].

\bibitem{NiauAkmansoy:2017kbw}
P.~Niau Akmansoy and L.~G.~Medeiros,
Eur. Phys. J. C \textbf{78}, no.2, 143 (2018)
[arXiv:1712.05486 [hep-ph]].

\bibitem{NiauAkmansoy:2018ilv}
P.~Niau Akmansoy and L.~G.~Medeiros,
Phys. Rev. D \textbf{99}, no.11, 115005 (2019)
[arXiv:1809.01296 [hep-ph]].

\bibitem{Ejlli:2020yhk}
A.~Ejlli, F.~Della Valle, U.~Gastaldi, G.~Messineo, R.~Pengo, G.~Ruoso and G.~Zavattini,
Phys. Rept. \textbf{871}, 1-74 (2020)
[arXiv:2005.12913 [physics.optics]].

\bibitem{Fedotov:2022ely}
A.~Fedotov, A.~Ilderton, F.~Karbstein, B.~King, D.~Seipt, H.~Taya and G.~Torgrimsson,
Phys. Rept. \textbf{1010}, 1-138 (2023)
[arXiv:2203.00019 [hep-ph]].

\bibitem{Ferraro:2007ec}
R.~Ferraro,
Phys. Rev. Lett. \textbf{99}, 230401 (2007)
[arXiv:0710.3552 [hep-th]].

\bibitem{Dauge:2012pw}
M.~Dauge and N.~Raymond,
J. Math. Phys. \textbf{53}, no.12, 123529 (2012)
[arXiv:1201.6471 [math.AP]].

\bibitem{Smirnov:2015gew}
Y.~G.~Smirnov and D.~V.~Valovik,
Phys. Rev. A \textbf{91}, 013840 (2015)

\bibitem{Zhou:2024pte}
R.~Zhou, M.~L.~N.~Chen, X.~Shi, Y.~Ren, Z.~Yu, Y.~Tian, Y.~Liu and H.~Lin,
IEEE Trans. Antennas Propag. \textbf{72}, no.2, 2058 (2024)
[arXiv:2401.00855 [physics.app-ph]].

\bibitem{Noble:2008aei}
A.~Noble,
Phys. Lett. A \textbf{372}, no.14, 2346-2349 (2008)

\bibitem{Mohaghegh:2019mpp}
M.~Mohaghegh, B.~A.~Arand and M.~Shahabadi,
Photonics Electromagn. Res. Symp. - Fall, PIERS - Fall - Proc., Xiamen, China, 2019, pp 858-865

\bibitem{Zhang:2020scp}
T.~Zhang, G.~Wang and D.~Deng,
Results Phys. \textbf{19}, 103625 (2020)

\bibitem{Zhou:2021ewp}
D.~Zhou, Y.~Ming, J.~Wang and C.~Gan,
Physica Scripta \textbf{96}, no.4, 045605 (2021)


\bibitem{Iukhtanov:2023rsw}
N.~G.~Iukhtanov and M.~V.~Rybin,
Opt. Lett. \textbf{48}, no.11, 3043-3046 (2023)


\bibitem{Sheremet:2021krw}
A.~S.~Sheremet, M.~I.~Petrov, I.~V.~Iorsh, A.~V.~Poshakinskiy and A.~N.~Poddubny,
Rev. Mod. Phys. \textbf{95}, no.1, 015002 (2023)
[arXiv:2103.06824 [quant-ph]].

\bibitem{Ashida:2021rls}
Y.~Ashida, T.~Yokota, A.~Imamoglu and E.~Demler,
Phys. Rev. Res. \textbf{4}, no.2, 023194 (2022)
[arXiv:2105.08833 [quant-ph]].

\bibitem{Ferraro:2003ia}
R.~Ferraro,
Phys. Lett. A \textbf{325}, 134-138 (2004)
[arXiv:hep-th/0309185 [hep-th]].

\bibitem{Manojlovic:2020ndn}
N.~Manojlovic, V.~Perlick and R.~Potting,
Annals Phys. \textbf{422}, 168303 (2020)
[arXiv:2006.09053 [math-ph]].

\bibitem{ATLAS:2017fur}
M.~Aaboud \textit{et al.} [ATLAS],
Nature Phys. \textbf{13}, no.9, 852-858 (2017)
[arXiv:1702.01625 [hep-ex]].


\bibitem{Hoshimov:2025tdx}
H.~Hoshimov, A.~Davlataliev, F.~Atamurotov, A.~Abdujabbarov and A.~{\"O}vg{\"u}n,
JHEAp \textbf{45}, 306-315 (2025)
\end{thebibliography}
\end{document}